# Suppression and revival of superconducting phase coherence in monolayer FeSe/SrTiO$_3$

H. Ru[1], Z. J. Li[1], S. Y. Wang[1], B. K. Xiang[1], Y. H. Wang[1,2*]

1. State Key Laboratory of Surface Physics and Department of Physics, Fudan University, Shanghai 200433, China
2. Shanghai Research Center for Quantum Sciences, Shanghai 201315, China

* To whom correspondence and requests for materials should be addressed. Email: wangyhv@fudan.edu.cn

**Monolayer FeSe grown on SrTiO$_3$ (FeSe/STO) is an interfacial high temperature superconductor distinctively different from bulk FeSe. Due to the fragility of this two-dimensional system in the atmosphere, the investigation of its intrinsic superconductivity and intertwined orders has largely been limited to surface-sensitive charge probes compatible with ultra-high vacuum environment. However, the superconducting phase coherence of the interface is challenging to probe. Here, we perform *in-situ* mutual inductance in ultra-high vacuum on FeSe/STO in combination with band mapping by angle-resolved photoemission spectroscopy (ARPES). We find that even though the monolayer showed a gap-closing temperature above 50 K, surprisingly no diamagnetism is visible down to 5 K. This is the case for few-layer FeSe/STO until it exceeds a critical number of 5 layers where diamagnetism suddenly appears. But the superfluid density does not saturate down to the base temperature in these thick samples. On the other hand, the suppression of diamagnetism in the few-layer FeSe/STO can be lifted by depositing a FeTe layer on top. The superconducting transition is much sharper than that in the thick FeSe/STO. However, $T_c$ and superfluid density both decrease with increasing FeTe thickness. Shining ultraviolet light on the FeTe/FeSe/STO heterostructure enhances $T_c$ similarly independent of the FeSe thickness, showing that the diamagnetism originates at the FeSe/STO interface.**

**Our observation may be understood by a scenario in which interfacial superconducting phase coherence is highly anisotropic.**

Phase coherence in a two-dimensional (2D) superconductor is a nontrivial part of the order parameter that distinguishes it from the bulk. While electron pairing and phase coherence are synonymous in a bulk Bardeen-Cooper-Schrieffer (BCS) superconductor [1], phase fluctuations in 2D due to vortex excitations limit the transition temperature $T_c$ below the pairing temperature (the famed Berezinskii-Kosterlitz-Thouless, or BKT transition [2,3]). In high-$T_c$ superconductors represented by the cuprate and iron-based families, ubiquitous emergent electronic and spin ordering often intertwine with superconducting phase coherence [4]. High-$T_c$ superconductivity in the 2D limit is even more enigmatic given the enhanced phase fluctuations [5,6]. What's more, the interfacial coupling from substrate or capping layer is capable of shifting the balance of any competition among charge, spin and phase.

Monolayer FeSe grown on $SrTiO_3$ (STO) is a particularly interesting case of 2D high-$T_c$ superconductor. It exhibited much enhanced pairing gap than that of bulk FeSe [7–9]. *In-situ* electrical transport measurements reported inhomogeneous superconducting transition with the resistance down-turn at $T_c^{onset}$ ~ 40 K and zero-resistance at $T_{c0}$ ~ 29 K [10], which were less than its pairing temperature of 65 K but well above the $T_c = 9$ K of bulk FeSe. Meanwhile, the highest $T_{c0}$ so far reported from *ex-situ* transport measurements was ~ 23 K as a FeTe capping layer was necessary for such measurement [11]. Curiously for multilayer FeSe/STO, the pairing gap was absent [12–14] and instead extensive electronic nematicity was present on the surfaces [14–17]. The charge nematicity was known to be affected by electrical doping [14], lattice strain and crystal anisotropy [16]. Diamagnetism was reported on a Co-doped multi-layer with Se capping [18].

In spite of these great progress since its original discovery, the relationship between Cooper pairing, charge nematicity and superconducting phase coherence in the mono- and multi-layer FeSe/STO remains unclear. Since the FeTe layer may affect the intrinsic superconductivity of FeSe/STO [19,20], it is important to directly measure Meissner diamagnetism on pristine monolayer FeSe/STO, for which the band-structure has been well determined. Furthermore, as the spectroscopic tools cannot access the buried interface in the multilayer FeSe/STO, its potential interfacial superconductivity is yet to be established and correlated with the prevalent charge orderings observed on the surface of the multilayers. However, such experiments are so far hampered by the lack of *in-situ* magnetic probes which are directly sensitive to the weak diamagnetic susceptibility of 2D superconductors.

In this work, we present *in-situ* two-coil mutual inductance study on mono- and multi-layer FeSe/STO to investigate their intrinsic superconducting phase coherence. This mutual inductance setup is inside an ultra-high-vacuum system where the samples are epitaxially grown and characterized by angle-resolved photoemission spectroscopy (ARPES). We show that even though the bare monolayers exhibit superconducting gap up to 65 K, their Meissner diamagnetism are suppressed down to 4.9 K. Diamagnetism occurs in the multilayers only above a critical thickness $d_c = 5$ unit-cells (UC). Meanwhile, deposition of 2UC FeTe on monolayer FeSe/STO immediately revives the diamagnetism. Increasing FeTe thickness reduces the $T_c$ and superfluid density for the mono- and multi-layers below $d_c$. These observations suggest a phase nematic order in pristine monolayer FeSe/STO.

We first characterize the band-structures of the mono- and multi-layers. We have grown FeSe films of well-controlled thickness $d$ (UC) using molecular-beam-epitaxy on conductive 0.5% Nb doped STO for ARPES measurements. We follow the similar growth and annealing procedures, which are described in detail in the Supplemental Material [21], as reported earlier [9]. The monolayer ($d =1$) FeSe/STO exhibits a hole band 60 meV below $E_F$ at the $\Gamma$ point and an electronic band with back-bending at the Fermi surface around the $M$ point (Fig. 1a). The symmetrized energy distribution curves at the Fermi momentum exhibit coherence peaks with finite pairing gaps below 50 K (Fig. 1b). The gap is 12.3 meV at 13 K and it reduces with increasing temperature following a BCS form (Fig. 1c). Such a band-structure and pairing gaps are in quantitative agreement with earlier works [8,22].

The band-structures are distinctively different in the multi-layers. The hole band at $\Gamma$ shifts up in energy to cross the Fermi level and the circular electron pocket at $M$ develops into a cross-like structure (Fig. 1d). Neither bands shows observable pairing gaps as in agreement with earlier reports [9]. The cross structure in films with $d \geq 2$ has been interpreted as a hallmark of charge nematicity because it recovers a circular shape through electron doping [23–25]. The consistency of the ARPES spectra of our mono- and multi- layers with those in earlier works attests to layer-by-layer deposition and the quality of our films. The lattice mismatch between FeSe and STO is also evident from the Fermi surface maps. The in-plane lattice constant $a$ of the surface FeSe layers can be determined from the size of the first Brillouin zone. $a$ is the largest in the monolayer at 3.9 Å but relaxes quickly with increasing $d$ towards the bulk value of 3.76 Å (Fig. 1d, green dashed line). But the relaxation is incomplete in the thickest samples we study. The presence of strain could explain that thick FeSe/STO does not show similar bulk superconductivity as in crystalline FeSe.

In order to obtain signature of intrinsic superconductivity of FeSe/STO, we perform *in-situ* mutual inductance measurements, which are very sensitive to weak diamagnetic inductance of superconducting thin films [18,26]. The technical advantage of magnetic measurement over charge transport for characterizing 2D superconductors is that it is contactless and does not require well-connected superconducting regions. Our two-coil coaxial mutual inductance device (Fig. 2a) is mounted inside the ARPES-MBE system so as to achieve *in-situ* growth and measurement. An alternat-current current ($I$) applied to the drive coil induces an inductive voltage ($V_p$) in the pick-up coil. The Meissner screening of the magnetic field from *I* by a superconducting film changes $V_p$. We calibrate the sensitivity of this setup by a 10-nm NbN film (Fig. 1a and SOM).

Unexpected for a superconductor with a pairing gap, there is no observable Meissner signal on either component of $V_p$ down to the base temperature of 4.9 K in the FeSe/STO monolayer (Figs. 2a and b). Diamagnetism is also absent in the multilayers with $d < 5$. But upon further increasing $d$, the in-phase component shows a broad peak whereas the out-of-phase component decreases with temperature below $T_c$. These features are consistent with the diamagnetic response of a conventional superconducting thin film such as NbN. The broadness of the peak in out-of-phase component of $V_p$ of FeSe/STO suggests that their phase transitions are highly inhomogeneous. These diamagnetic features in bare multi-layers are quite different from the mutual inductance signal from a multi-layer with Co doping and Se capping [18]. There, the out-of-phase component decreased below 65 K and saturated around 10 K. But no peak was observed from the in-phase component. The magnetic dopants and surface strain may influence the charge nematicity in the bare multi-layers [14] [16] and alter the magnetic response.

By converting the in-and-out of phase components of $V_p = X + iY$ into a complex surface impedance $Z_s = R_s + i\omega L$ [27–30], we obtain the Pearl length Λ [21], which characterizes the penetration depth in a 2D superconductor. The inverse of the Pearl length ($Λ^{-1}$) is proportional to the sheet superfluid density $n_s$ by only universal constants (Fig. 2c). $Λ^{-1}$ of multilayered FeSe/STO are not only more than 10 times smaller than that of NbN with a similar thickness, their temperature dependence is also completely different. It increases with reducing temperature in a sub-linear fashion akin to that of overdoped cuprates [31]. None of the superconducting multilayers reaches saturation in superfluid density at our base temperature. We have measured more than 18 films with various $d$ and $d_c = 5$ appears to be the critical thickness at which superfluid density can be detected (Fig. 2d). The strain on the diamagnetic side is below 2% but there is no clear correlation between the $T_c$ and the strain (Fig. 2d, right axis) as determined from the ARPES maps (Fig. 1d). We recall that previous *in-situ* electrical transport showed zero-resistance below 29 K in pristine monolayer FeSe/STO [10]. The absence of diamagnetism for $d < d_c$ strongly suggests that phase coherence is suppressed at the interface below an intermediate length scale, which is determined by the inverse of local $n_s$.

To understand the origin of the suppression of superconducting phase coherence, we epitaxially grew FeTe over the mono- and multi-layer FeSe/STO. As-grown FeTe without oxidization is non-superconducting [29]. We show representative samples above and below $d_c$ ($d = 1$ and $d = 12$) for comparison. For the monolayer, diamagnetism immediately emerges after depositing 2 layers of FeTe. It exhibits a transition temperature of 18 K (Fig. 3a). Surprisingly, with increasing FeTe thickness *t*, both $T_c$ and superfluid density decrease monotonously. Few layer samples less than $d_c$ show similar revival of diamagnetism with 2 layers of FeTe and reduction of both $T_c$ and superfluid density upon further increasing *t* (Fig. S6). Such reduction with increasing *t* suggests that the revival effect introduced by the FeTe layers are not from filling any holes in the FeSe layer. For $d = 12$, which is already diamagnetic without FeTe, finite *t* does not increase $T_c$ but does enhance the superfluid density with increasing *t* (Fig. 3b). The zero-resistance temperature obtained from charge transport (*ex-situ*) on these two samples with the thickest *t* (Figs. 3c and d) are consistent with the $T_c$'s determined from the mutual inductance measurements.

The effect of lattice strain from FeTe on the FeSe/STO interfacial superconductivity is worth considering first. Strain is known to enhance the monolayer's pairing gap [32,33]. And our observation of reduced $T_c$ and superfluid density with increasing *t* is consistent with weakened pairing as the strain on the monolayer is partially relaxed by thicker FeTe. However, such an effect from the strain is only secondary as the revival of phase coherence cannot be explained by it. The reduction of strain caused by FeTe (bulk $a = 3.82$ Å) is less than that of the same amount of

FeSe (bulk $a = 3.76$ Å). Therefore, two layers of FeTe should have less structural effect on the monolayer than two layers of FeSe, which is still less than $d_c$ and the phase coherence should still have been suppressed (Fig. 2d). The sudden appearance of diamagnetism with $t = 2$ of FeTe on the monolayer suggests that the revival of interfacial phase coherence has a non-structural origin.

Since charge correlation is prevalent in FeSe/STO, we use UV excitation to investigate its relationship with superconducting phase coherence. The UV light (365 nm) penetrates the top FeTe and FeSe layers and reaches the STO substrate. It induces metastable polar distortion in STO [19] which strongly affects the charge accumulation at the FeSe/STO interface. We find that it enhances $T_c$ of the monolayer with $t = 48$ by about 2.8 K and enhances the superfluid density as well (Fig. 3e). It has a very similar effect on the $d = 12$ sample as seen by the temperature dependence of increased superfluid density (Fig. 3f). The effects brought by UV excitation can be removed by cycling the temperature to 300 K. These observations are indication that carrier concentration directly controls the interfacial superconducting phase coherence.

In a 2D superconductor, there is a linear dependence between the superfluid density and $T_c$ at a BKT transition. Earlier works showed that ultrathin cuprates in the under-doped regime and intercalated FeSe exhibited close-to-linear dependence [31,34–36]. By extracting the superfluid density from the mutual inductance at the base temperature for various $t$, we obtain the variation of superfluid density as a function of $T_c$ (Fig. 4a). There is again a clear distinction above and below $d_c$. For $d > d_c$, $T_c$ barely changes with increasing superfluid density after the slight increase due to the first two FeTe layers. But for $d < d_c$, the group of points for each $d$ can be well-fit with a line going through the origin. This difference further shows that the superconductivity in the thick multilayers has a different origin from that in the mono- and few-layers. The linear dependence between superfluid density and $T_c$ in the latter provides evidence for a BKT transition at $T_c$. However, the superfluid density does not exhibit a sharp jump at $T_c$ (Fig. 3a) as is expected for BKT of a large and homogeneous 2D superconductor [35]. Noting that the jump of superfluid density for a BKT transition can be smeared out due to finite-size effect [5,6] or disorder, we can infer that the revived interfacial phase coherence by the FeTe layers is spatially inhomogeneous.

The completely suppressed phase coherence in superconducting monolayer FeSe/STO further suggests the order parameter is not only inhomogeneous but also highly anisotropic. Zero-resistance in charge transport and Meissner effect are the defining macroscopic signatures of a uniform superconductor and they occur simultaneously [1]. The observations of the pairing gap and zero-resistance measured on the pristine monolayer FeSe/STO seem to contradict our observation of the absence of Meissner effect. The contradiction can be resolved when anisotropy is considered. Dissipationless charge transport only requires phase coherence along a certain direction, whereas Meissner screening of magnetic field needs circulating supercurrent and thus isotropic phase coherence. Based on the above observations and considerations, we propose a picture that reconciles the disparity in the charge and phase aspects of superconductivity: a nematic order where phase of the complex superconducting order parameter is uniform within the domain but randomly different between the domains (Fig. 4b).

We emphasize that such a phase nematic order at the FeSe/STO interface is distinct from the charge anisotropies observed by scanning tunneling microscopy [15]. Firstly, such electronic nematicity has been reported on the surface of the multilayers but not on the monolayers. Secondly, as such charge ordering occurs at the surface where pairing is absent, it is unclear whether it will compete with the interfacial superconductivity as in the case of electronic nematicity in bulk iron-arsenides [37,38]. Nevertheless, the phase nematicity may be induced by the strong charge correlation at the interface which we have shown earlier (Figs. 3e and f). Assuming that is true, the spatial inhomogeneity of superfluid density after the phase coherence is revived by FeTe or UV excitation is consistent with vestigial charge ordering [39].

In conclusion, we have observed by *in-situ* mutual inductance that superconducting phase coherence in pristine mono- and few-layer FeSe/STO is suppressed. It is revived at a $T_c$ of 18 K by depositing two layers of FeTe. The dependence of the revival on FeSe/FeTe thickness and UV excitation provides clear evidence that the superconductivity originates at the FeSe/STO interface. When the FeSe is below a critical thickness of 5 unit-cells, $T_c$ and superfluid density follow a linear relationship, further suggesting the two-dimensional nature of the superconducting phase transition. Our observations hint at a phase nematic state in pristine monolayer FeSe/STO due to charge correlation at the interface, which calls for further investigation.

## Acknowledgement

YHW would like to acknowledge support by National Natural Science Foundation of China (Grant No. 11827805 and 12150003), National Key R&D Program of China (Grant No. 2021YFA1400100) and Shanghai Municipal Science and Technology Major Project (Grant No. 2019SHZDZX01). All the authors are grateful for the experimental assistance by T. Zhang and R. Peng and for the stimulating discussions with D. L. Feng, J. Zhao and L. Shu.

## Data availability

The data that support the findings of this work are available from the corresponding author upon reasonable request.

*Materials: Nematicity and Beyond*, Annu. Rev. Condens. Matter Phys. **10**, 133 (2019).

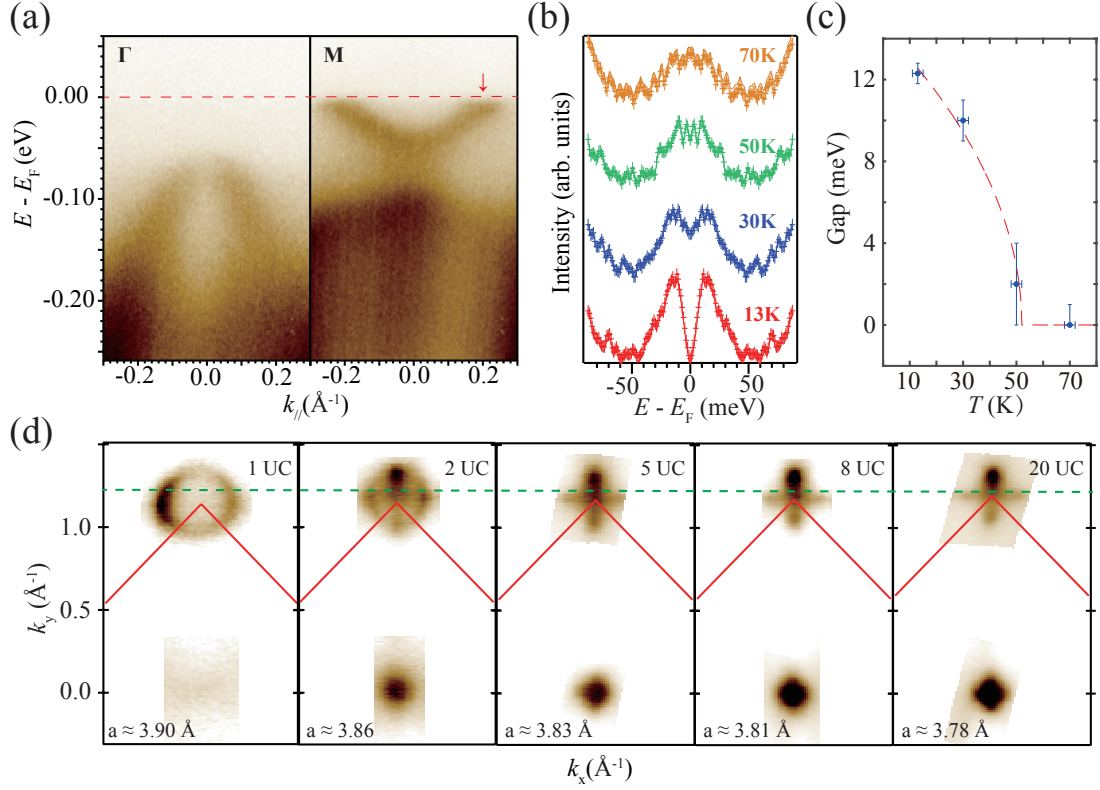

**Fig. 1. Angle-resolved photoemission spectra of mono- and multi-layer FeSe/SrTiO$_3$.** (a) Energy-momentum spectra of monolayer FeSe/SrTiO$_3$ (STO) at Γ (left) and M (right) respectively measured at 13 K. (b) Temperature dependence of the energy distribution curves at the Fermi momentum (the red arrow in (a)) symmetrized about the Fermi energy ($E_F$). (c) The superconducting gaps as a function of temperature $T$ extracted from the data in (b). The red dashed line is a fitting of the gap-temperature dependence based on the BCS theory. (d) Fermi surface mapping of the mono- and multi-layer FeSe at 13 K. The number of FeSe layers (*d*) in unit-cell (UC) is labeled on each panel. The intensity is integrated over a 10 meV window from $E_F$. The red solid lines mark the Brillouin zone boundary. Green dashed lines represent the position of $k_y$ at the M point of bulk FeSe ($a = 3.76$ Å). Superconducting gaps are absent in the multilayer FeSe/STO.

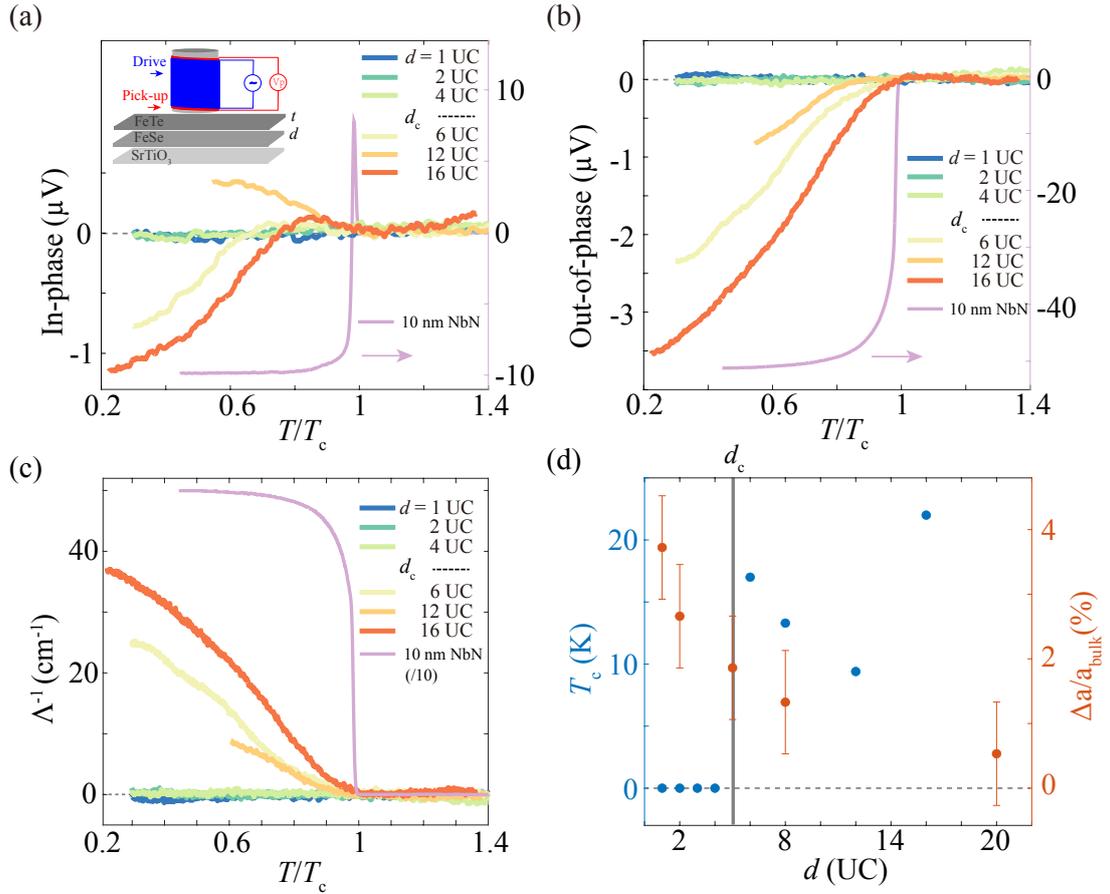

**Fig. 2. Intrinsic suppression of superconducting phase coherence in bare monolayer FeSe/STO.** (a) and (b) are in-phase and out-of-phase components, respectively, of the inductive voltage ($V_p$) as a function of reduced temperature ($T/T_C$, where $T_C$ is the critical temperature). Data are acquired on bare FeSe/STO (without FeTe on top) with various thickness $d$. Right axis: $V_p$ on a 10-nm NbN film used as a calibration for penetration depth. (a) inset: schematic of the mutual inductance measurement performed *in-situ* in ultra-high vacuum to detect the Meissner effect. Diamagnetism is absent below a critical FeSe thickness $d_c = 5$ UC. (c) Inverse Pearl length $\Lambda^{-1}$ (proportional to the sheet superfluid density $n_s$) as a function of reduced temperature extracted from (a) and (b). (d) Left axis: $T_C$ versus $d$ obtained from the mutual inductance signal. Right axis: strain versus $d$ obtained from the Fermi surface maps in Fig. 1d. The presence of superconducting gap but the absence of Meissner diamagnetism suggest that the superconducting phase coherence is intrinsically suppressed in monolayer FeSe/STO.

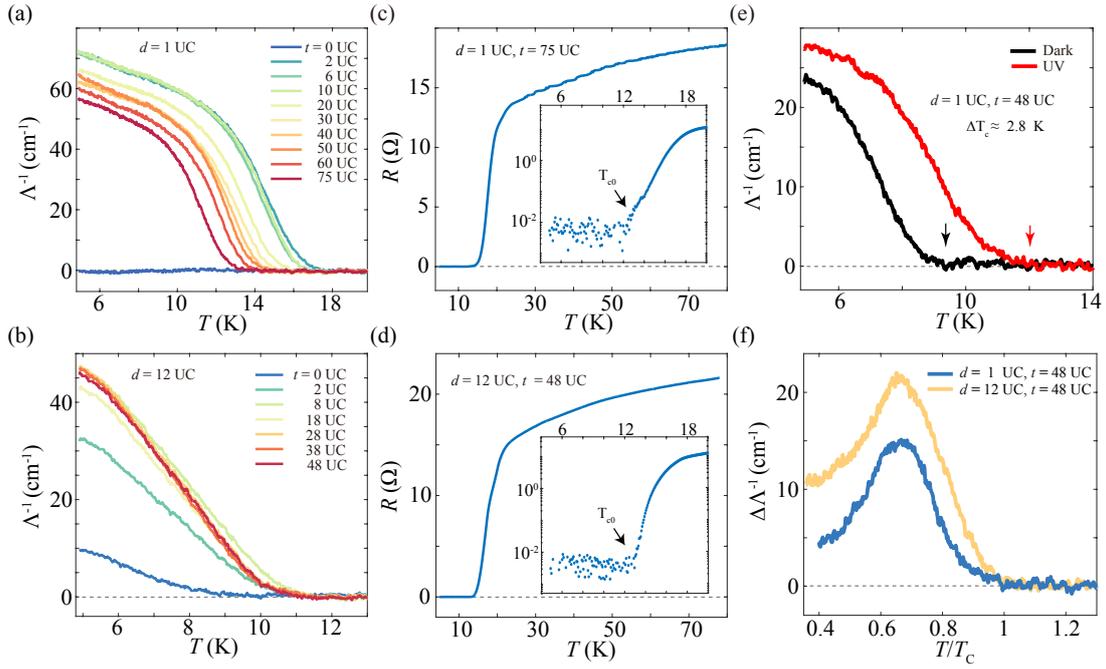

**Fig. 3. Revival of FeSe/STO interfacial phase coherence by FeTe deposition.** (a) and (b) $\Lambda^{-1}$ of a monolayer ($d = 1$) and a multilayer ($d = 12$) FeSe/STO, respectively, with different thickness of FeTe ($t$) deposited on top. (c) and (d), Resistance as a function of temperature for the monolayer and the multilayer FeSe/STO, respectively, after the growth sequence of FeTe. (e) $\Lambda^{-1}$ before (black line) and after (red line) ultraviolet (UV) light exposure for FeTe($t = 48$)/FeSe($d = 1$)/STO. (f) The difference of $\Lambda^{-1}$ for two films of FeTe($t = 48$)/FeSe($d = 1$)/STO (blue line) and FeTe($t = 48$)/FeSe($d = 12$)/STO (orange line) before and after UV light exposure ($\Delta\Lambda^{-1}$). The sample in (e) is not the same as in (c), whereas (f) ($d = 12$) and (b) are the same one.

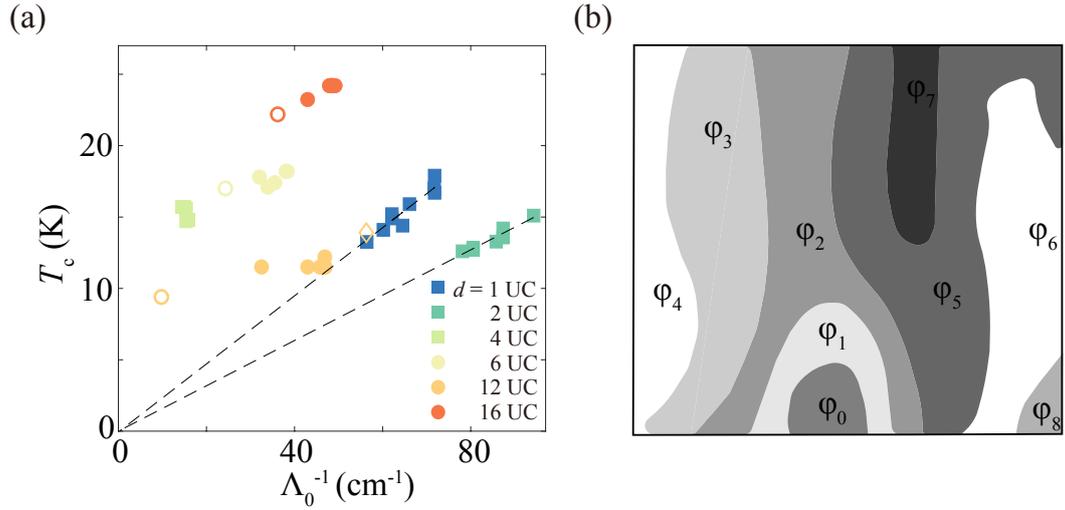

**Fig. 4. Linear dependence of superfluid density on $T_c$.** (a) The relationship between $T_c$ and $\Lambda_0^{-1}$ (at 4.9 K) extracted from the temperature curves of mutual inductance measured on various $t$ of FeTe over FeSe($d$)/STO. The hollow circles represent bare FeSe($d$)/STO, the diamond point is the film ($d = 12$, $t = 48$) after UV exposure. The linear relationship for $d < d_c$ suggests a two-dimensional superconducting phase transition. (b) A schematic of speculated phase nematicity in monolayer FeSe/STO. Different color zones represent nematic phase domains with random relative superconducting phase in between.